\documentclass[twocolumn,nofootinbib]{revtex4-1}
\usepackage{latexsym,epsfig,amssymb, amsmath,nicefrac}

\usepackage{amsfonts,amsthm} 
\usepackage{bm,bbm}
\usepackage{graphicx}
\usepackage{esint}
\usepackage{color}
\usepackage{braket}
\def\K{K{\"a}hler}

\newcommand{\be}{\begin{equation}}
\newcommand{\ee}{\end{equation}}
\newcommand{\bea}{\begin{eqnarray}}
\newcommand{\eea}{\end{eqnarray}}
\def\bs{\begin{subequations}}
\def\es{\end{subequations}}

\newcommand{\rf}[1]{(\ref{#1})}

\begin{document}

\title{{\Large More on Universal Superconformal Attractors}}

\author{Renata Kallosh}

\affiliation{{}Department of Physics and SITP, Stanford University, \\ 
Stanford, California 94305 USA, kallosh@stanford.edu}

\begin{abstract}
We define a general class of superconformal inflationary attractor models \cite{Kallosh:2013tua,Kallosh:2014ona}, which include, among others,  inflationary models of  `induced gravity' which were argued to retain perturbative unitarity up to the Planck scale \cite{Giudice:2014toa}.  The choice of the function $f(\phi)$ in superconformal attractors for the models introduced in \cite{Giudice:2014toa} is $f(\phi)= \phi^n - \xi^{-1}$. We present both superconformal and supergravity versions of these models, which were derived in  \cite{Kallosh:2013tua,Kallosh:2014ona} for arbitrary $f(\phi)$, together with the universal conditions required for stabilization of the extra three moduli present in the superconformal attractors, in addition to the inflaton.
\end{abstract}

\maketitle

\smallskip

The  universal  inflationary attractors  \cite{Kallosh:2013tua,Kallosh:2014ona} are defined by one arbitrary function of a single real scalar field, $\Omega(\phi)$. This class of models in Jordan frame can be represented as follows:
 \begin{align}
 {1\over \sqrt {-g }}\, \mathcal{L}_{\rm J} =   \tfrac12  \Omega(\phi) R  - \tfrac12 (\partial \phi)^2 -  {\lambda^2\over \xi^2}  (\Omega (\phi)- 1) ^2  \, . \label{Baction} 
\end{align}
In  the  Einstein frame this model is 
 \begin{align}
 {1\over \sqrt {-g }}\,  \mathcal{L}_{\rm E} &=      \tfrac12   R - \tfrac12 \Big (\Omega(\phi)^{-1} + \tfrac32 (\log \Omega(\phi))'^2\Big ) (\partial \phi)^2 \nonumber \\
 &- \frac{{\lambda^2}  (\Omega (\phi)- 1)^2}{ \xi^2\Omega(\phi)^2}    \,. \label{Einstein}
 \end{align}
 The superconformal generalization of these models is defined by the \K\, potential of the embedding space and the superpotential, which both depend on an arbitrary holomorphic function $\Omega(\Phi)$  \cite{Kallosh:2014ona}.  The bosonic model in \rf{Baction}  and \rf{Einstein} follows from the superconformal model   \cite{Kallosh:2014ona} and from the corresponding supergravity model presented in \cite{Kallosh:2013tua}, upon stabilization of the extra non-inflaton scalars. The corresponding  stability analysis was performed  in \cite{Kallosh:2013tua} for an arbitrary function $\Omega(\Phi)$, and is valid for any choice of $\Omega(\Phi)$, as long as the slow-roll parameters during inflation are small. We will describe below the universal supersymmetric attractor models of \cite{Kallosh:2013tua,Kallosh:2014ona}. As we will see, this simultaneously provides a superconformal and supergravity embedding of the bosonic model studied in  \cite{Giudice:2014toa}.

A class of models\footnote{The class of models presented in  \cite{Giudice:2014toa} was argued to be unitarity-safe. 
The issue of perturbative unitarity violation is somewhat controversial, see, for example, a discussion in
\cite{Ferrara:2010in}. Here we will not discuss it and just show that all models of induced gravity studied in \cite{Giudice:2014toa}  can be embedded in the general class of the universal superconformal attractor models of  \cite{Kallosh:2013tua,Kallosh:2014ona}.} called `induced gravity' studied in \cite{Giudice:2014toa} corresponds to a choice
\be
\Omega_{\rm ind}(\phi)= \xi f_{\rm ind}(\phi) \ .
\ee
Here $f_{\rm ind}(\phi)$ is what was called $f(\phi)$ in \cite{Giudice:2014toa}; we will keep the notation $f(\phi)$ for the function introduced in \cite{Kallosh:2013tua,Kallosh:2014ona}.
In refs. \cite{Kallosh:2013tua,Kallosh:2014ona}  the Jordan  frame model \rf{Baction}  was introduced  for an arbitrary function $\Omega(\phi)$ and was shown to lead to an Einstein frame action \rf{Einstein}.  
Particular examples of $\Omega (\phi)$ studied in \cite{Kallosh:2013tua,Kallosh:2014ona} were of the form
\begin{align}
 \Omega(\phi)  = 1 + \xi f(\phi)\, ,  
\label{O}\end{align}
where $f^{2}(\phi)$ was a $\xi$-independent polynomial function $\phi^{2n}$, or $1+\cos(\phi/c)$.  In this way our superconformal examples in  \cite{Kallosh:2013tua,Kallosh:2014ona} allowed us to study simple supersymmetric  
models interpolating between a large set of chaotic inflation models (or natural inflation models) at $\xi=0$ and the corresponding attractors points in the $(n_s, r)$-plane for increasing values of $\xi$, representing a non-minimal coupling to gravity $\xi f(\phi) R$. However,  the general superconformal/supergravity construction developed in \cite{Kallosh:2013tua,Kallosh:2014ona} allows arbitrary $\Omega (\phi)$ and therefore arbitrary dependence of $f(\phi)$ on $\xi$.

To see how all induced gravity  models studied in \cite{Giudice:2014toa} are described in the context of general models developed in \cite{Kallosh:2013tua,Kallosh:2014ona} defined in eqs. \rf{Baction} and \rf{Einstein},  we have to identify the functions $\Omega_{\rm ind}(\phi)$ with $ \Omega(\phi)$  
\begin{align}
   \Omega_{\rm ind}(\phi)=  \Omega(\phi) \, \qquad \Rightarrow \qquad        \xi f_{\rm ind}(\phi)= 1 + \xi f(\phi) \ ,
\end{align}  
which shows that 
\be
 f(\phi)=      f_{\rm ind}(\phi) - \xi^{-1} \ .
\label{rel}\ee
Examples of induced gravity with $ f_{\rm ind}(\phi)=\phi^n$ studied in \cite{Giudice:2014toa} are given by our equations \rf{Baction}, \rf{Einstein} and \rf{O} under condition that
\be
 f (\phi)=      \phi^n - \xi^{-1}
 \label{compare}\ee
for all values of $n$. It was suggested in \cite{Giudice:2014toa} that 
 induced gravity models and universal attractor models belong to the same class of models only for $n=1$.\footnote{A related observation about the unitarity-safe universal attractor models with  $\Omega(\phi)  = 1 + \xi \phi$ was made earlier in \cite{Kehagias:2013mya}.}
 This is indeed the case in the simplest examples with $\xi$-independent functions $f(\phi)$ discussed in \cite{Kallosh:2013tua}. In such case our relation given in eq. \rf{compare} is not possible.  However, as we already explained, the superconformal and supergravity models underlying eqs. \rf{Baction} and \rf{Einstein} are defined for arbitrary function $\Omega(\phi)$. Therefore, the condition of $\xi$-independence of $f(\phi)$ is not necessary, which allows to compare these models by identifying their functions $\Omega(\phi)$ and leads to eq.  \rf{compare} for any $n$.
All induced gravity models described in sec. 5 of \cite{Giudice:2014toa} are defined in our equations \rf{Baction}, \rf{Einstein}, \rf{O}, \rf{rel}, which explains the embedding of \cite{Giudice:2014toa} in the more general class of models introduced in \cite{Kallosh:2013tua,Kallosh:2014ona}.


A generic superconformal version of the 
models introduced in \cite{Kallosh:2013tua,Kallosh:2014ona} is
defined by the  \K\, potential of the embedding space, see \cite{Kallosh:2014ona}:
   \bea
&\mathcal{N}(X,\bar X)= -{1\over 2} |X^0|^2 \left[   \Omega (\phi)
+  \Omega(\bar \phi)\right] + |\Phi|^2 + |S|^2 \nonumber \\
&- {1\over 12}  |X^0|^2  \left[ \phi^{2}+\bar \phi^{2}\right] - 3 \zeta {(S\bar S)^2\over |X^0|^2 \Big[\Omega (\phi) + \Omega (\bar \phi) \Big]} 
 \,  \, ,
\label{general}
 \eea
where we are using the following notation for the complex field which has a zero conformal weight under local Weyl transformations:
$
\phi\equiv \sqrt{6} \,  { \Phi\over X^0} 
$.
Here $\Omega(\phi) $  is a holomorphic function,
and the superpotential is
\be
{\cal W}=  {\lambda\over \xi} \,  {(X^0)^2\over 3} \, S  \,  (\Omega(\phi)- 1)\, .
\ee
This model in the Einstein frame leads to  supergravity with $X^0= \sqrt 3$ and $\phi=\sqrt{2}\,  \Phi $ defined by the following \K\ potential and superpotential \cite{Kallosh:2013tua,Kallosh:2014ona}:
 \begin{align}
\hskip -10pt K =  & - 3 \log[ \tfrac12 ( \Omega(\phi) + \Omega( \bar \phi))   - \tfrac13 S \bar S + {1\over 12} (\phi - \bar \phi)^2 \notag \\
& +  \zeta \frac{( S \bar S)^2}{\Omega(\phi) + \Omega( \bar \phi)}] \,, \qquad   
  W = {\lambda\over \xi}  S \, (\Omega(\phi)- 1)  \, .
\label{KW2} \end{align}
This leads exactly to the bosonic model \rf{Baction} discussed in  \cite{Kallosh:2013tua}  upon identifying the real part of the $\Phi$-field with the inflaton, $\Phi = \bar \Phi= \phi / \sqrt{2}$, while $S=0$, 
which is a consistent truncation. During inflation the masses of the non-inflaton fields are
$m_{{\rm Im}\, \Phi}^2  = ( 4/3 + 2 \epsilon - \eta ) V$,  $m_S^2  = ( - 2/3 + 6 \zeta + \epsilon ) V$, where $\epsilon$ and  $\eta$ are the slow-roll parameter, computed in case of arbitrary functions $\Omega$ \cite{Kallosh:2013tua}.
Up to slow-roll corrections, with the choice $\zeta > 1/9$, all non-inflaton fields are heavy, they have masses $m^2 > H^2$ and quickly reach their minima at $S=\Phi-\bar \Phi=0$, meanwhile the inflaton $\phi$ is light, $ m_{\phi}^2  = \eta V\ll V$.

The minimum of the potential is a supersymmetric extremum with  $DW=W=0$ at $S=\Phi-\bar \Phi=0$. At this minimum, in the class of models  with $f(\phi)=  f_{\rm ind}(\phi) - \xi^{-1}$ one should have
\be
\Omega(\phi) = \Omega( \bar \phi) =1\,  \quad \Rightarrow  \quad f(\phi)=  f_{\rm ind}(\phi) - \xi^{-1}=0 \ .
\ee
In such models  $\phi= \xi^{-1/n}$ in the supersymmetric minimum, in agreement with the property of a bosonic  model studied in \cite{Giudice:2014toa}.

In conclusion we would like to stress here that the superconformal attractors have an important universality property, first discovered in the models of conformal inflation in \cite{Kallosh:2013hoa}, where the values of $n_s$ and $r$ during inflation are independent on the choice of the potential $V(\varphi)$, which can be a rather  general  function $f^2(\tanh \varphi)$, for a canonical field $\varphi$. The analogous kind of universality we have seen in strong coupling attractors and $\alpha$-attractors in \cite{Kallosh:2013tua}. The models of induced gravity studied in \cite{Giudice:2014toa} also exhibit an attractor behavior since $n_s$ and $r$ during inflation are independent on the choice of  $n$ in the function  $f_{\rm ind}(\phi)=\phi^n$. As we have shown here, these models can be easily embedded in the general class of superconformal attractor models \cite{Kallosh:2013tua,Kallosh:2014ona} presented  here in  eqs.\rf{Baction}, \rf{Einstein}, \rf{O}.
This immediately provides the bosonic induced gravity models studied in \cite{Giudice:2014toa} with a consistent supersymmetric generalization presented above. It is based on the stability analysis performed for generic superconformal attractors with an arbitrary function $\Omega(\phi)$ in \cite{Kallosh:2013tua}, which valid for all models of this class where slow-roll inflation is possible.

I am grateful to A. Linde and D. Roest  for collaboration on the work in \cite{Kallosh:2013tua} on which the current note is based, and for the useful discussions.   This work was supported by the NSF Grant PHY-1316699, SITP and the Templeton grant ``Quantum Gravity frontier''.

\providecommand{\href}[2]{#2}\begingroup\raggedright\endgroup

\end{document}